\documentclass{elsart}

\usepackage{graphicx}

\usepackage{amssymb}

\begin{document}

\begin{frontmatter}

\title{Return distributions in dog-flea model revisited}
\author{Burhan Bakar}
\address{Department of Physics, Faculty of Science, Ege University, 35100 Izmir, Turkey}

\author{Ugur Tirnakli\corauthref{cor}}
\corauth[cor]{Corresponding author.}
\ead{ugur.tirnakli@ege.edu.tr}
\address{Department of Physics, Faculty of Science, Ege University, 35100 Izmir, Turkey\\
and \\ 
Division of Statistical Mechanics and Complexity, 
Institute of Theoretical and Applied Physics (ITAP) Kaygiseki Mevkii, 
48740 Turunc, Mugla, Turkey}

\begin{abstract}
A recent study of coherent noise model for the system size independent case provides an 
exact relation between the exponent $\tau$ of avalanche size distribution and the $q$ 
value of the appropriate $q$-Gaussian that fits the return distribution of the model. This 
relation is applied to Ehrenfest's historical dog-flea model by treating the fluctuations 
around the thermal equilibrium as avalanches. We provide a clear numerical evidence that 
the relation between the exponent $\tau$ of fluctuation length distribution and the $q$ 
value of the appropriate $q$-Gaussian obeys this exact relation when the system size is large 
enough. This allows us to determine the value of $q$-parameter \emph{a priori} from 
one of the well known exponents of such dynamical systems. Furthermore, it is shown 
that the return distribution in dog-flea model gradually approaches to $q$-Gaussian as the 
system size increases and this tendency can be analyzed by a well defined analytical 
expression. 
\end{abstract}

\end{frontmatter}

\section{Introduction}

Since the pioneering work of Bak, Tang, and Wiesenfeld (BTW) in 1987 \cite{BakPRL59}, 
the self-organized criticality (SOC) paradigm has seen a burst of activity in the literature. 
In their seminal paper, by making use of so-called BTW sandpile model, these authors demonstrated 
that systems under their natural evolution are driven at a very slow rate until one of their 
elements reaches a critical stationary state and this triggers a sudden activity, \emph{i.e.}, 
avalanche. When the avalanche is over, the system evolves again according to the slow drive 
until a next avalanche is triggered without having any fine-tuning of parameters. 
Such a dynamic gives rise to the power-law correlations seen in the non-equilibrium steady 
states \cite{Jensen,Bak-Book}.

Following the BTW sandpile model a great variety of models ranging from the deterministic and 
stochastic to the dissipative and conservative ones have been introduced which exhibit the 
phenomenon of SOC (for an overview, see \cite{Dhar} and references therein). 
Among them, a random neighbor version of the BTW sandpile model shows that a dynamical system 
with only two degrees of freedom can exhibit SOC and the dynamics can be described by a master 
equation \cite{Flyvbjerg-PRL}. 
Soon after the random neighbor BTW sandpile model was introduced, its conservative variant 
has been reported \cite{Nagler-PRE,Nagler-JSP}. This conservative variant is neither 
extended nor dissipative with regard to the amount of sand in the system but still shows SOC with 
critical exponents where the dynamics of the model is given by a Fokker-Planck equation. 
The avalanche size distribution $P(\lambda)$ is readily obtained by solving the Fokker-Planck 
equation at an absorbing boundary and is shown to exhibit a power-law regime 
$P(\lambda)\sim\lambda^{-\tau}$, followed by an exponential tail. Indeed, this model is an 
adaptation of the famous dog-flea model introduced by Ehrenfest in 1907 \cite{Ehrenfest} to 
describe the process of approaching an equilibrium state in a large set of uncoupled two state 
systems in the presence of fluctuations (avalanches) around this state \cite{Kac1,Kac2}.

In a recent work, the SOC feature of the dog-flea model was studied by simulating the 
underlying stochastic process that describes the natural evolution of the model \cite
{BBUTPRE09}. In this paper, we show that the relation between the power-law 
exponent $\tau$ of the fluctuation length distribution and the $q$ value of the appropriate 
$q$-Gaussian that fits the return distribution (\emph{i.e.}, distribution of fluctuation length 
differences at subsequent time steps) obeys the rule
\begin{equation}\label{eq:q-tau}
q=e^{1.19\tau^{-0.795}},
\end{equation}       
which was reported for the limited number of earthquakes from the World and California 
catalogs \cite{CarusoPRE75}. This approximate relation (\ref{eq:q-tau}) enables one to 
obtain the value of $q$ parameter \emph{a priori} from the power-law exponent $\tau$ of 
fluctuation length distribution. Using the same line of thought, the $\tau$ value for the 
dog-flea model was obtained as $\tau\simeq1.517$ by making use of the maximum 
likelihood estimation method yielding to a value of $q=2.35$ of the concomitant $q$-Gaussian 
that fits the return distribution at $N\rightarrow\infty$ limit \cite{BBUTPRE09}. 

Soon after the relation (\ref{eq:q-tau}) was reported, an exact relation between 
$\tau$ and $q$ has been introduced for the coherent noise model (CNM) in the size 
independent case as \cite{ACUTCNM}, 
\begin{equation}\label{eq:q-tau2}
q=\frac{\tau+2}{\tau}.
\end{equation} 
These relations between $\tau$ and $q$ given by Eqs. (\ref{eq:q-tau}) and (\ref{eq:q-tau2}) are 
slightly different from one another. Both relations approach $q=1$ at $\tau\rightarrow\infty$ limit 
while only the latter achieves the value $q=3$ for the given $\tau=1$ value 
\cite{ACUTCNM}. Let us also remark that if we identify $\tau = 1/(q_{s} -1)$, we can obtain 
from Eq.~(\ref{eq:q-tau2}) $q_{s} -1 = (q-1)/2$, which is precisely Eq.~(30) in Ref.\cite{alvarez}.   

The chosen system size plays a crucial role in analyzing the SOC feature of the model. 
At relatively small system sizes, the distribution of the fluctuation length time-series 
$\lambda(t)$ in dog-flea model exhibits a power-law regime, which is followed by an 
exponential decay because of the finite size effects, while at large system sizes, 
the power-law regime increases and the exponential decay is postponed \cite{BBUTPRE09}. 
This size dependent behavior of the fluctuation length distributions 
results in different topological properties of the return distributions 
as we discuss in subsequent sections.              

Our main task will be to analyze the SOC in the dog-flea model through numerical evaluation 
of the underlying stochastic process. In this manner the fluctuation length and return 
distributions are studied when the system size is large enough, \emph{i.e.}, at the 
$N\rightarrow\infty$ limit, and for small system sizes (\emph{i.e.}, when the finite size 
effects are visible). The rest of the paper is organized as follows. 
The model and the numerical procedure that we implement are given in the next section. 
Then, the probability distribution of fluctuation length for the system size, chosen so that 
the finite size effects are avoided, is obtained by numerical evaluations. 
Once the power-law exponent $\tau$ of the fluctuation length distribution is found, 
the distribution of returns are analyzed. 
Next, we analyze the return distribution of the dog-flea model when the finite size effects are visible. 
A summary and discussion of the results conclude this communication.

\section{Dog-flea model}

The dynamics of the dog-flea model has simple rules. The model has $N$ dynamical sites 
represented by the total number of fleas shared by two dogs (dog $A$ and dog $B$). 
Suppose that there are $N_{A}$ fleas on dog $A$ and $N_{B}$ fleas on dog $B$ 
leading to a population of fleas $N=N_{A}+N_{B}$. For convenience, $N$ is assumed 
to be even. In every time step, a randomly chosen flea jumps from one dog to the other. 
Thus, we have $N_{A}\rightarrow N_{A}\pm1$ and $N_{B}\rightarrow N_{B}\mp1$. 
The procedure is repeated for an arbitrary number of times. In the long time run, the mean 
number of fleas on both dog $A$ and dog $B$ converges to the equilibrium value, 
$\langle N_{A}\rangle =\langle N_{B}\rangle = N/2$ with the fluctuations around these 
mean values. A single fluctuation is described as a process that starts once the number of 
fleas on one of the dogs becomes larger (or smaller) than the equilibrium value $N/2$ and 
stops when it gets back to the value $N/2$ for the first time. Thus, the end of one 
fluctuation specifies the start of the subsequent one. The length ($\lambda$) of a fluctuation 
is determined by the number of time steps elapsed until the fluctuation ends.

\begin{figure*}[t]
\begin{center}
\includegraphics[width=9cm]{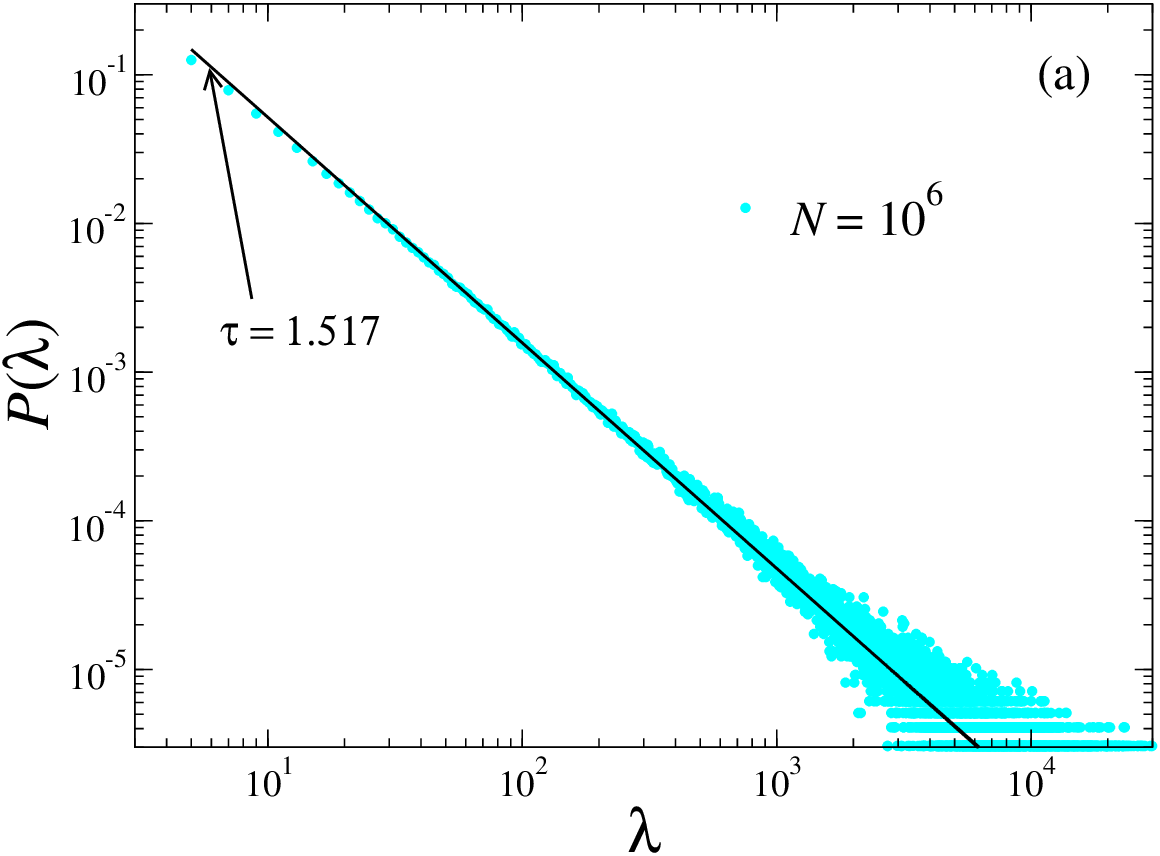}
\includegraphics[width=9cm]{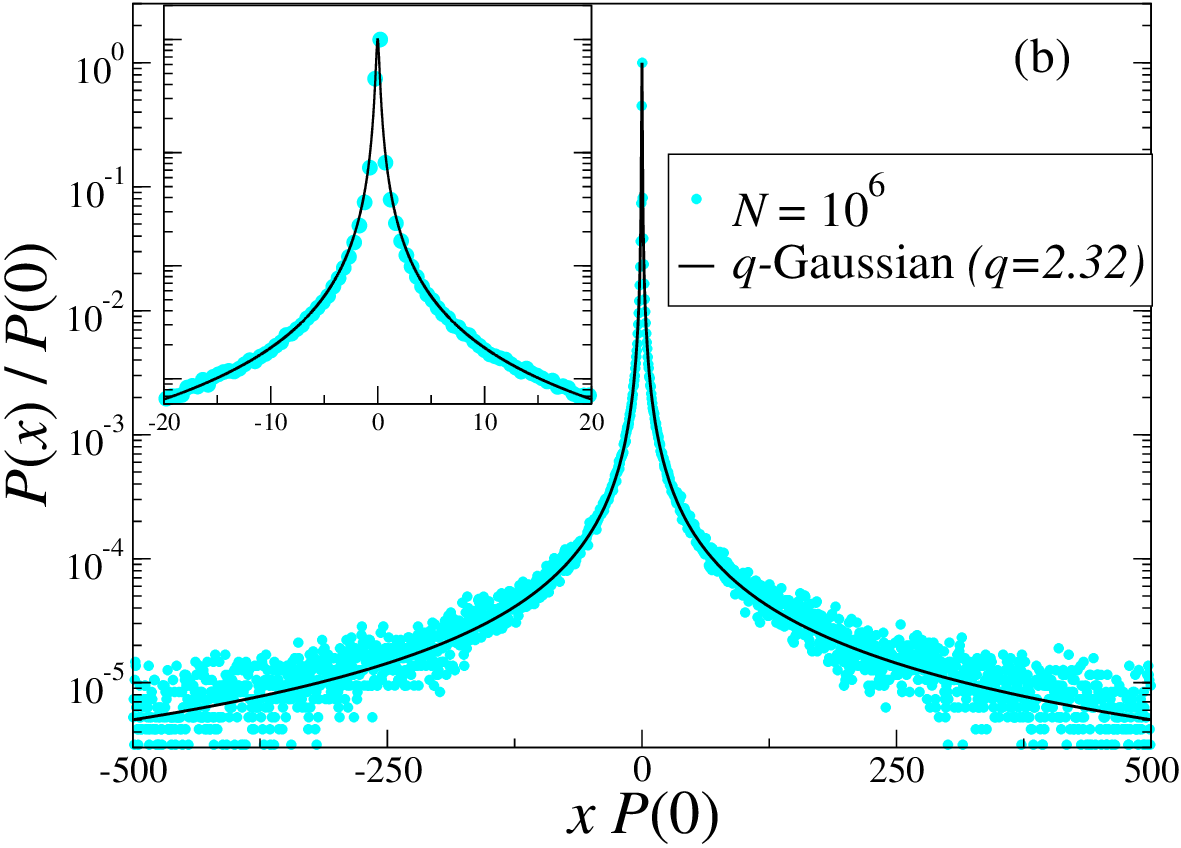}
\end{center}
\caption{(Color online) (a) Fluctuation length distribution for $N=10^{6}$. The full 
black line represents the fitting curve of the distribution with slope $\tau\simeq1.517$. 
The distribution has an arbitrary normalization such that $P(\lambda=1)=1$. (b) Return 
distribution for $N=10^{6}$. The full black curve represents the appropriate $q$-Gaussian with 
$q=2.32$ determined from relation (\ref{eq:q-tau2}) \emph{a priori} and 
$\beta=30$. The central part of the distribution is emphasized in the inset.}
\label{fig:fig1}
\end{figure*}
\section{Fluctuation length and return distributions}
In order to study the behavior of return distributions with large $N$ the 
system size is chosen as $N=10^{6}$, enabling the finite size effects 
to appear only at very large fluctuation lengths. 
In Fig.~\ref{fig:fig1}(a), we plot the distribution of the fluctuation length time-series 
$\lambda(t)$ obtained from $2\times10^{9}$ fluctuations. 
It can then be seen that the fluctuation length distribution follows a power-law regime, 
$P(\lambda)\sim\lambda^{\tau}$ with $\tau\simeq1.517$ and exponential decay (\emph{i.e.}, 
finite size effects) is not visible.  

In Fig.~\ref{fig:fig1}(b), the distribution of returns, \emph{i.e.}, the difference between 
fluctuation lengths obtained at consecutive time steps, as $\Delta\lambda(t)=\lambda(t+1)-
\lambda(t)$ is shown. At this point it should be emphasized that in order to have zero 
mean, the returns are normalized by introducing the variable $x$ as $x=\Delta\lambda-
\langle\Delta\lambda\rangle$, where $\langle\cdots\rangle$ stands for the mean value of the 
given data set. From Fig.~\ref{fig:fig1}(b), it is clear that the signal of return distribution 
for $N=10^{6}$ is not Gaussian. Instead, it exhibits fat tails which is described by 
$q$-Gaussian,
\begin{equation}\label{eq:q-gauss}
P(x)=P(0)[(1+\bar{\beta}(q-1)x^{2}]^{1/(1-q)}, 
\end{equation}
where $\bar{\beta}$ characterizes the width of the distribution and $q$ is the index of 
nonextensive statistical mechanics \cite{Tsallis88,Tsallis}. As suggested in \cite{ACUTCNM}, the 
parameter $q$ can be determined directly from the relation (\ref{eq:q-tau2}) \emph{a priori} as 
$q=2.32$. In Fig.~\ref{fig:fig1}(b), the solid black line represents the appropriate $q$-Gaussian 
with $q=2.32$ and $\beta=30$ that perfectly fits the signal of return distribution not only for 
the tails but for the intermediate and the very central part (see inset). 
\begin{figure*}[t]
\begin{center}
\includegraphics[width=9cm]{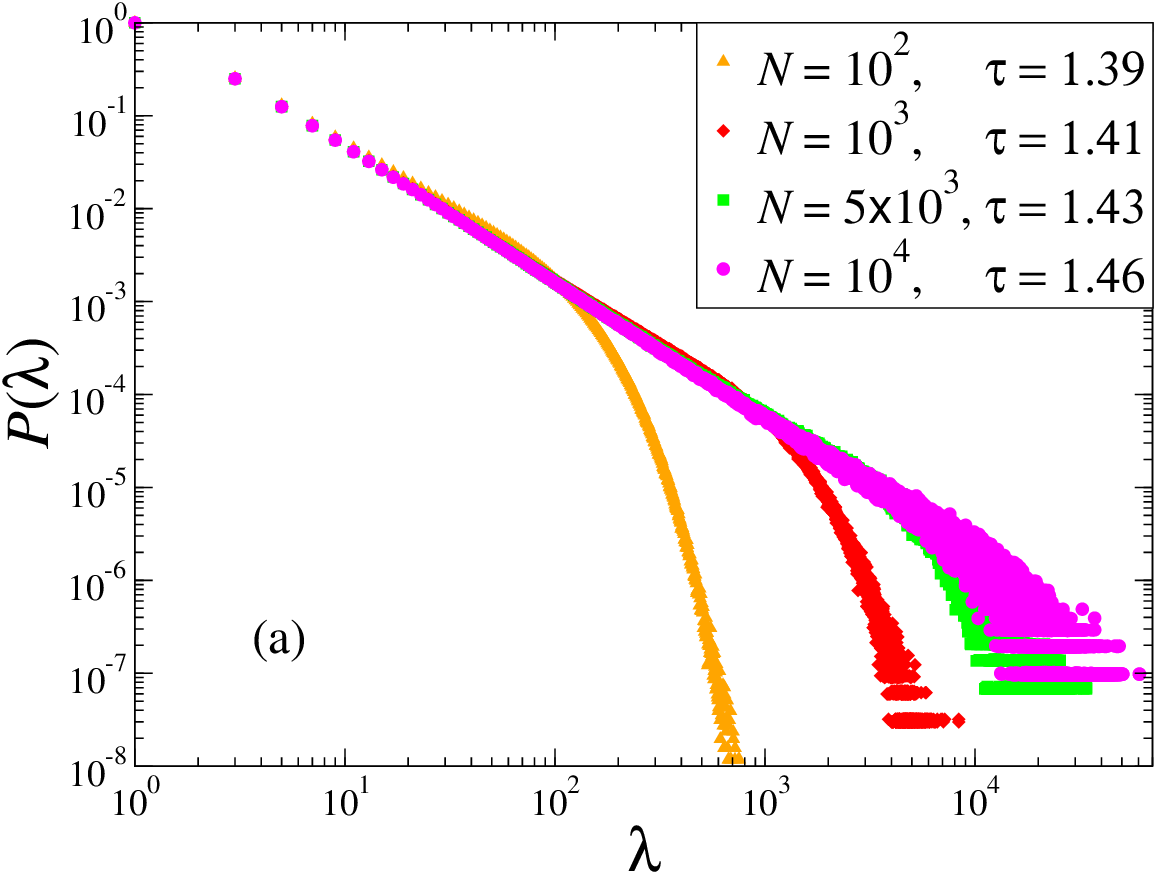}
\includegraphics[width=9cm]{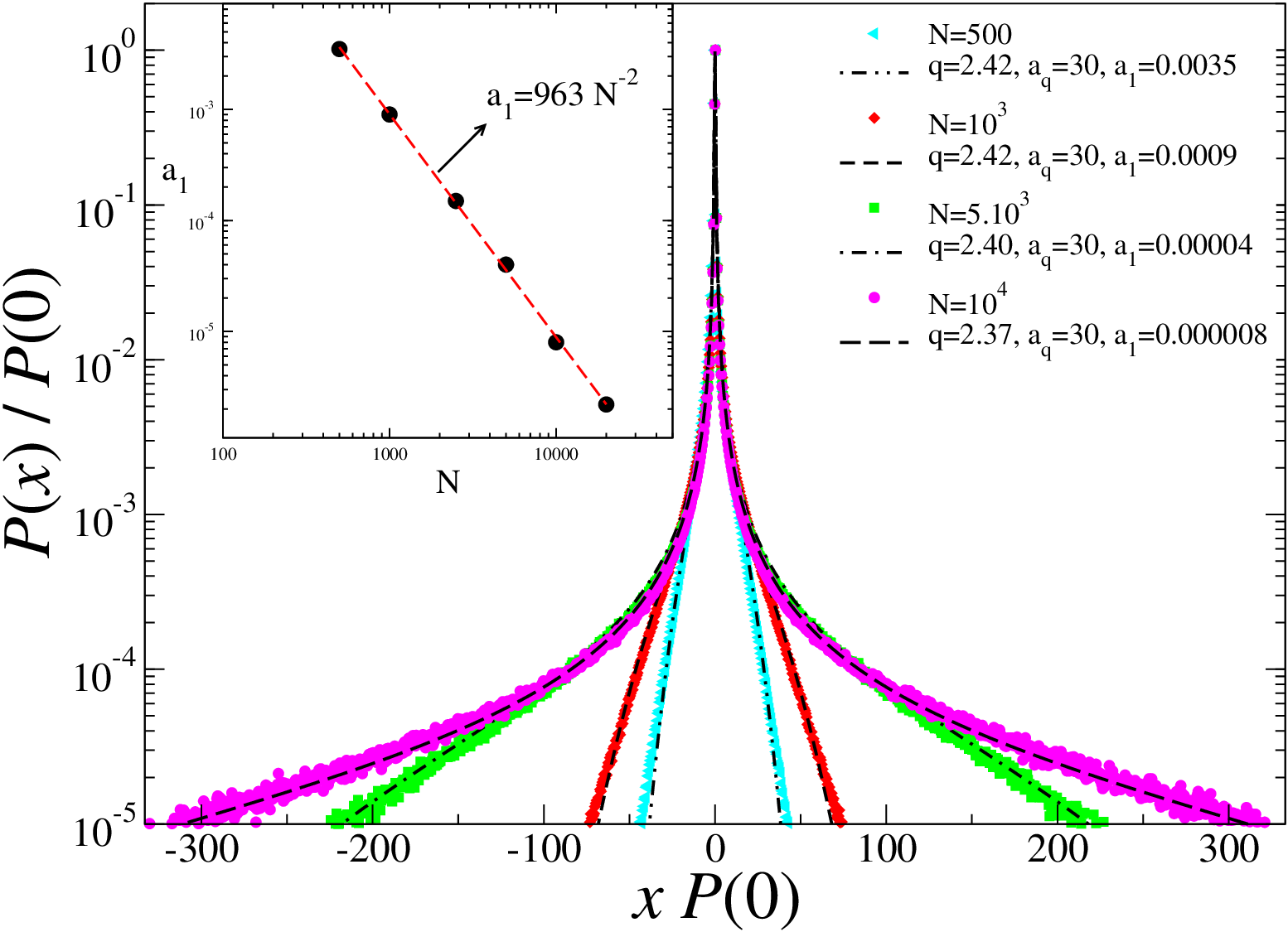}
\end{center}
\caption{(Color online) (a) Fluctuation length distributions for $N=10^{2}$, 
$N=10^{3}$, $N=5\times10^{3}$ and $N=10^{4}$. $10^{9}$ fluctuations are considered for 
each different system sizes. The legends of figure give the corresponding slope values 
$\tau$ of power-law regimes. The distributions have an arbitrary normalization such that 
$P(\lambda=1)=1$. 
(b) Return distributions for $N=10^{2}$, $N=10^{3}$, $N=5\times10^{3}$ and $N=10^{4}$. 
The black lines on each return distribution represent 
the fitting curves obtained by Eq.~(\ref{eq:semiq}) with $q$ values 
determined \emph{a priori} from the relation (\ref{eq:q-tau2}) using the corresponding 
$\tau$ values.  For each curve $a_{q}=30$ which is equal to $\beta$ value of appropriate 
$q$-Gaussian that fits the return distribution with largest system size, \emph{i.e.}, 
$N=10^{6}$. In the inset, we also give the parameter $a_1$ as a function of $N$ in order 
to better visualize how the finite-size effect disappears as $N$ increases. It is clear 
that $a_1$ appears to vanish as $a_1=963 N^{-2}$.}
\label{fig:fig2}
\end{figure*}

The fluctuation length distribution $P(\lambda)$ in dog-flea model shows size dependent 
behavior. Its signal exhibits a power-law regime following by an exponential regime 
because of finite-size effect at relatively small system sizes. However, as it has been 
reported in \cite{BBUTPRE09}, the finite-size scaling hypothesis is satisfied. In Fig.~\ref
{fig:fig2}(a), the fluctuation length distribution of $10^{9}$ fluctuations for four 
different system sizes $N=10^{2}$, $N=10^{3}$, $N=5\times10^{3}$ and $N=10^
{4}$ are presented. As the system size increases the power-law regime gets longer and the 
exponential decay is postponed. Since the system size is far away from being at the $N
\rightarrow\infty$ limit (\emph{i.e.}, not large enough to avoid the finite size effects), each 
distribution of fluctuation length has different power-law slopes with slightly different $\tau$ 
values. This difference between the $\tau$ values of each distribution of fluctuations for 
different system sizes leads to return distributions, which are neither Gaussian nor 
$q$-Gaussian. Nevertheless, from Fig.~\ref{fig:fig2}(b) it is obvious that as the system size $N$ 
increases, leading to a longer power-law regime in the fluctuation length distribution, the 
curves of return distributions start to exhibit a tendency to a kind of fat tailed 
distribution, \emph{i.e.}, $q$-Gaussian. This gradual approach to $q$-Gaussian in the 
presence of finite-size effect is described by the following distribution \cite
{Tsallis,TsallisPLA,TsallisUT}, 
\begin{equation}\label{eq:semiq}
y=\left[1-\frac{a_{q}}{a_{1}}+\frac{a_{q}}{a_{1}}e^{(q-1)a_{1}x^{2}}\right]^{1/
(1-q)}. 
\end{equation}
If $a_{1}=0$, Eq.~(\ref{eq:semiq}) coincides with the $q$-Gaussian and a Gaussian is 
recovered for $a_{q}=a_{1}$ (\emph{i.e.}, $q=1$).  

As it is the case in CNM \cite{ACUTCNM}, Eq.~(\ref{eq:semiq}) seems to coincide 
with the signal of the return distribution when the finite-size effects are invisible in dog-flea 
model. As it is seen from Fig.~\ref{fig:fig2}(b), each return distribution corresponding to 
different system sizes exhibits a topological behavior which is fitted by Eq.~(\ref{eq:semiq}) 
with appropriate $q$ values determined \emph{a priori} from the relation (\ref{eq:q-tau2}). 
Each critical exponent $\tau$ for different system sizes is obtained by considering only 
the power-law regime of corresponding fluctuation length distribution, 
see Fig.~\ref{fig:fig2}(a). Fig.~\ref{fig:fig2}(b) also reveals the fact that the longer 
the power-law regime persists for fluctuation length distribution, the better the appropriate 
$q$-Gaussian dominates in the return distribution. Eventually, as $N\rightarrow\infty$, 
the power-law regime of fluctuation length distribution is expected to continue forever 
(see Fig.~\ref{fig:fig1}(a)). It is also evident from the inset of Fig.~\ref{fig:fig2}(b) that 
the finite-size effects vanish as $N^{-2}$, which is completely the same tendency found 
in \cite{ACUTCNM} for the coherent noise model. 
When the finite-size effects disappear, the corresponding return distribution seems to 
converge to the $q$-Gaussian for the entire region (see Fig.~\ref{fig:fig1}(b)).

\section{Conclusion}

We analyze the SOC in Ehrenfest's dog-flea model through the probability distributions of 
fluctuation length and of the differences between the fluctuation lengths at subsequent time 
steps, \emph{i.e.}, returns, by simulating the stochastic process that describes the evolution of 
model. The fluctuations around the thermal equilibrium are treated as avalanches. In order to avoid 
the finite-size effects the size of the system is chosen large enough, \emph{i.e.,} $N=10^{6}$. 
This enables one to obtain a power-law regime with slope $\tau\simeq1.517$ without any exponential 
decay. Then, the signal of return distribution is analyzed and it is shown that it converges to 
a $q$-Gaussian with $q=2.32$, a value obtained \emph{a priori} from Eq.~(\ref{eq:q-tau2}). This $q$ 
value is slightly different from the one reported in \cite{BBUTPRE09}, where $q=2.35$ has been obtained 
using Eq.~(\ref{eq:q-tau}). The difference between the $q$ values for the same distribution stems 
from the fact that Eq.~(\ref{eq:q-tau}) is an approximate relation slightly differing from 
the exact relation given by Eq.~(\ref{eq:q-tau2}) (see \cite{ACUTCNM} for details). 

The case where the finite-size effects are visible is also investigated by extensive simulations. 
When the system size is chosen relatively small, the finite-size effects appear in the 
avalanche size distribution: the distribution follows a power-law regime with corresponding slope 
value $\tau$ followed by an exponential decay. In this case, the return distributions numerically 
converge to appropriate $q$-Gaussian starting from the central part and gradually evolves towards the 
tails as the system size increases. This behavior of return distribution is indeed in good 
agreement with the one obtained when the system size is large enough. 
It is also found that the finite-size effects vanish as $N^{-2}$. 
As $N\rightarrow\infty$ the signal of return distributions for the finite system sizes 
numerically converges to a $q$-Gaussian in the entire region.

\section*{Acknowlegment}
This work has been supported by TUBITAK (Turkish Agency) under the Research Project number 104T148 
and by Ege University under the Research Project number 2009FEN027.

\bibliographystyle{elsart-num}
\bibliography{FleasPRL}
\end{document}